\documentclass[10-pt]{article}

\begin{document}

\title{Electromagnetic mass-models in general relativity
reexamined}
\author{J. Ponce de Leon\thanks{E-Mail:
jponce@upracd.upr.clu.edu}  \\Laboratory of Theoretical Physics, 
Department of Physics\\ 
University of Puerto Rico, P.O. Box 23343,  
San Juan,\\ PR 00931, USA}
\date{October 2003}

\maketitle
\begin{abstract}
The problem of constructing a model of an extended charged particle 
within the context of general relativity  
 has  a long and distinguished history. 
The distinctive feature of these models is that, in some 
way or another,  they require 
the presence of 
negative mass in order to maintain stability against Coulomb's 
repulsion. Typically, the particle contains a 
core of $negative$ mass surrounded by a positive-mass outer layer,
 which 
emerges from the Reissner-Nordstr\"{o}m field.
 In this work  we
show how the Einstein-Maxwell field equations can be used to
 construct an extended model where the mass is positive  
 everywhere. This requires the principal pressures to  
be unequal inside the
 particle. The model is obtained by setting the ``effective" matter 
density, rather than the rest matter density, equal to zero. The 
Schwarzschild mass of the particle arises from the electrical and 
gravitational field (Weyl tensor) energy. The model satisfies the 
energy conditions of Hawking and Ellis. A particular solution that 
illustrates the results is presented.  

\end{abstract}

\section{Introduction} 
A point charge is incompatible with classical electrodynamics,
 because it leads  to the well-known self-energy and
 stability problems as well as to the occurrence of ``runaway" 
solutions of the Lorentz-Dirac equations. 
One 
way to overcome these problems is to assume that charged particles
are built as singularity-free concentrations of fields that, 
however small, have finite size. 

The first model of an ``extended" charged particle was studied by
 Abraham \cite{Rohrlich}. 
This model ascribed the $entire$ mass of the particle 
to the interaction energy with its own electromagnetic field.
This would make the particle radius $R$ equal to
\begin{equation}
\label{radius}
R = \frac{Q^{2}}{M c^{2}},
\end{equation}
where $Q$ and $M$ are the charge and mass of the particle. This 
quantity is commonly known as the ``classical electron 
radius" \cite{Schweber}.

Shortly after,  it was realized that this model was unstable 
and  inconsistent with the Lorentz transformations of special relativity. A mechanism to 
overcome the electrostatic repulsion was 
suggested by Poincar$\acute{e}$. He postulated the existence of  
non-electromagnetic cohesive forces that would hold the charge 
together, and make the model compatible with special 
relativity. Since
 these forces provide a phenomenological, 
rather than a fundamental, description of the particle, 
this mechanism is not really satisfactory.

Today, the Abraham-Lorentz-Poincar$\acute{e}$ model
 for an extended charge belongs to the history of physics. And
 the problems associated with the point charge theory 
are
 overcome in quantum electrodynamics via renormalization, without 
the necessity of introducing 
extended particles.  

However, this does not mean that the interest in the description 
 of extended charged particles  has been lost. And although 
charged particles obviously 
belong to the quantum
 domain, today it is 
understood that the concept of particle structure does 
not negate the notion of ``elementarity" \cite{CR}. Rather, 
extended particles are still in use to model the actual 
particle structure and extract relevant physical 
predictions.  

In general relativity extended models have been used 
by several authors to discuss some important aspects in the theory.
 For example, the role of 
gravitation 
in charged-particle formation has been 
analyzed by Cooperstock and Rosen 
 \cite{CR}. The
 relevance of the equation of state of ``false vacuum''
 $\rho = -p$ to relativistic electromagnetic mass models has been 
discussed by Gr{\o}n \cite{Gron} and by Tiwari, Rao and 
Kanakamedala \cite{TRK}. The phenomenon of gravitational repulsion
 around elementary 
particles like electrons has been investigated by a number of authors
\cite{CR}-\cite{PdL1}. Also, the validity of singularity 
theorems inside the electron has been  discussed by Bonnor and Cooperstock 
\cite{BC} as well as by the present author \cite{PdL2}.

All  the models for extended charged 
particles, used in the literature, exhibit the following ``peculiar"  
feature:
They need the presence of some  {\em negative} mass 
to maintain stability against Coulomb's repulsion.
 That is, independently of the working 
assumptions of the specific model, the picture of a (classical) charged particle is 
always the same: 
the particle should consist of a 
core of $negative$ mass surrounded by a positive-mass outer layer,
 which 
emerges from the Reissner-Nordstr\"{o}m field.

However, in ``conventional" physics, the mass is always positive.  
And, although one can invoke that macroscopic physics does not 
hold within charged particles, it is natural to ask whether it is 
possible or not to avoid the use of negative masses in the 
structure of charged particles. That is, without the introduction 
of a negative 
mass, can one construct an extended model
for a charged particle?

The object of this work is to show that the answer to this question is 
positive. In Sec. 2, we show how the Einstein-Maxwell equations can be 
used to construct a model of a charged particle whose gravitational 
and inertial masses are nowhere negative. In Sec. 3, we discuss the 
condition for the mass to be of electromagnetic origin. In Sec. 4, we give a 
simple example that illustrates the fact that the model satisfies 
the 
energy
conditions of Hawking and Ellis. Sec. 5 is a summary and discussion.    

\section{Structure of the Abraham-Lorentz-Poincar$\acute{e}$ particle}

 In its rest frame, the charge will be described by a static, 
spherically symmetric distribution of matter, which is 
assumed to be  governed by the 
Einstein-Maxwell equations. 

 We choose the line element 
in curvature coordinates  
\begin{equation}
\label{metric}
ds^{2} = e^{\nu}dt^{2} - e^{\lambda}dr^{2} - r^{2}(d\theta^{2} + 
sin^{2}\theta d\phi^{2}),
\end{equation}
 where $\nu$ and $\lambda$ are functions of $r$ alone. 
In these coordinates
 the energy-momentum tensor $T_{\mu\nu}$ is diagonal, viz.,
\begin{equation}
\label{EM tensor}
T^{\mu}_{\nu} = diag\;(M^{0}_{0} + \frac{E^{2}}{8\pi},\;\; M^{1}_{1} + \frac{E^{2}}
{8\pi},\;\; M^{2}_{2} - \frac{E^{2}}{8\pi}, \;\; M^{3}_{3} - \frac{E^{2}}{8\pi}),
\end{equation}
where $(0,1, 2, 3) \equiv (t, r, \theta, \phi),$ E is the usual electric field
intensity, $M_{\mu\nu}$ represents the energy-momentum tensor associated with the
``matter" contribution, and $M^{2}_{2} = M^{3}_{3}$ because of the spherical 
symmetry (We note that the symmetry {\em does not} require 
 $M^{1}_{1} = M^{2}_{2}$).

The electrovacuum region around the particle is described by 
the Reissner-Nordstr\"{o}m field, which,
 in curvature
 coordinates, has the form\footnote{In what follows
we use gravitational units: $c = G = 1$.}
\begin{equation}
\label{RN}
ds^{2} = (1 - \frac{2M}{r} + \frac{Q^{2}}{r^{2}})dt^{2} - 
(1 - \frac{2M}{r} + \frac{Q^{2}}{r^{2}})^{-1} dr^{2} 
 - r^{2}(d\theta^{2} + 
sin^{2}\theta d\phi^{2}),
\end{equation}
The charge inside a sphere of radius $r$ is given by
\begin{equation}
\label{charge}
q(r) =  4\pi \int_{0}^{r}{\rho}_{e}r^{2}dr,
\end{equation}
where ${\rho}_{e}$
is the charge density\footnote{This quantity is related to the
 ``proper" charge density 
$\hat{\rho}_{e}$ by ${\rho}_{e} = e^{\lambda/2}\hat{\rho}_{e}$.}.
Therefore, the total charge is $Q \equiv q(R)$. 
 
The ``effective" gravitational
 mass inside a sphere of radius $r$ is given by the Tolman-Whittaker formula,
 viz.,
\begin{equation}
\label{TW integral}
M_{G}(r) = 4\pi\ \int_{0}^{r}(T^{0}_{0} - T^{1}_{1} - T^{2}_{2} - T^{3}_{3})r^{2}
e^{(\nu + \lambda)/2}dr
\end{equation}
By analogy with 
(\ref{charge}), the quantity
\begin{equation}
\label{effective density}
\mu(r) = [(T^{0}_{0} - T^{1}_{1} - T^{2}_{2} - T^{3}_{3})e^{(\nu + \lambda)/2}],
\end{equation}
 can 
be interpreted as an ``effective" gravitational mass density.

The total mass $M$ in (\ref{RN}) is then $M = M(\infty)$. That is 
\begin{equation}
\label{total mass}
M \equiv 4\pi\ \int_{0}^{R}\mu_{in}r^{2} dr + 
 4\pi\ \int_{R}^{\infty}\mu_{out}r^{2} dr,
\end{equation}
where the subscripts ``in" and ``out" mean inside and outside the 
particle, respectively. 

Outside the particle $M_{\mu\nu} = 0$, and $E^{2} = Q^{2}/r^{4}$.
Therefore, using (\ref{EM tensor}), (\ref{RN}) and 
(\ref{effective density}), we find 
  $\mu_{out}= Q^{2}/(4 \pi r^{4})$. Therefore, the second term in 
(\ref{total mass})
can integrated to get
\begin{equation}
\label{bare mass + elect mass}
M = 4\pi\ \int_{0}^{R}\mu_{in}r^{2} dr + \frac{Q^{2}}{R}.
\end{equation}
Now, in order to construct the relativistic version of the  
old Abraham-Lorentz-Poincar$\acute{e}$ model
 for an extended charge \cite{Rohrlich},
\cite{Schweber}, we set $R = Q^{2}/M$.  
Thus, from (\ref{bare mass + elect mass}) we find
\begin{equation}
\label{strong}
\int_{0}^{R}\mu_{in}r^{2} dr = 0.
\end{equation}
We now assume that the effective gravitational mass density is 
nowhere negative, viz.,
\begin{equation}
\label{non-negative density}
 \mu(r) \geq 0.
\end{equation}
Consequently, 
from (\ref{strong}) and 
(\ref{non-negative density}) it follows that  $\mu_{in}(r)$ 
must vanish everywhere within the source, viz., 
\begin{equation}
\label{vanishing density}
\mu_{in}(r) =  0.
\end{equation}
In general relativity, because of the $linear$ relation between 
the curvature tensor and $T_{\mu\nu}$, the strong energy condition requires 
$R_{\mu\nu}V^{\mu}V^{\nu} \geq 0$ for 
an arbitrary non-spacelike vector $V^{\mu}$. Therefore, our assumption 
(\ref{non-negative density})  
is equivalent to assuming  that the ``strong" energy condition 
is applicable within the 
particle. 

Let us now write the Einstein-Maxwell equations associated to 
(\ref{metric}) 
\begin{equation}
\label{density}
8\pi\rho + E^{2} = - e^{-\lambda}(\frac{1}{r^{2}}\ - \frac{{\lambda}'}
{r}) + \frac{1}{r^{2}},
\end{equation}
\begin{equation}
\label{radial pressure}
 - 8\pi p_{r} + E^{2} = -e^{-\lambda}(\frac{1}{r^{2}} + \frac{{\nu}'}{r}) +
\frac{1}{r^2},
\end{equation}
\begin{equation}
\label{tangential pressure}
 - 8\pi p_{\perp} - E^{2} = -\frac{e^{-\lambda}}{2}( {\nu}'' +
 \frac{{\nu'}^{2}}{2} + \frac{{{\nu}'}-{{\lambda}'}}{r} - 
\frac{{\nu}'{\lambda}'}{2}),
\end{equation}
\begin{equation}
\label{electric field}
 E(r) = \frac{q(r)}{r^{2}},
\end{equation}
where  $\rho \equiv M^{0}_{0}$, $p_{r}\equiv - M^{1}_{1}$ and
 $p_{\perp} \equiv -  M^{2}_{2} = - M^{3}_{3}$ denote the rest
 energy density
and the principal pressures of the matter present, respectively.
The primes denote differentiation with respect to $r$. In this
 notation the condition (\ref{vanishing density}) reduces to
\begin{equation}  
\label{strong energy condition}
 \mu_{in} = (\rho + p_{r} + 2p_{\perp} + \frac{E^{2}}{4\pi})e^{(\nu + \lambda)/2} = 0,
\end{equation}

\subsection{Unequal principal pressures}
 Let us immediately note that if $p_{r}$ were 
equal to $p_{\perp}$, then 
the particle would contain some negative ``rest (or inertial)"  
mass density. Indeed, at the boundary $r = R$, 
(\ref{strong energy condition}) would reduce 
to $\rho(R) = - Q^{2}/(4 \pi R^{4}) < 0$, because the 
continuity of $E$, $\nu$,
 $\lambda$, and  $ \nu'$, requires $E^{2}(R) = Q^{2}/R$ and
$p(R) = 0$.

 The conclusion, therefore, is that to construct an 
Abraham-Lorentz-Poincar$\acute{e}$ model
 for an extended charge with
 $(I)$ {\em Everywhere non-negative gravitational
 mass} and $(II)$ {\em Everywhere positive rest mass density}, 
 the particle 
{\em must have unequal principal pressures}. 

 \subsection{$M_{G} = 0$ inside the particle} 
From the field equations, we find that 
(\ref{strong energy condition}) is 
equivalent to
\begin{equation}
\label{strong condition}
(r^{2}e^{(\nu -\lambda)/2}\nu')' = 0.
\end{equation}
The regularity conditions as well as the condition of local 
flatness at the center demand 
$\nu' \rightarrow 0$ as $r \rightarrow 0$. 
Therefore, from (\ref{strong condition}) 
it follows that $\nu' = 0$ and $M_{G} = 0$ 
throughout the source (although the ``inertial"  
mass  $4 \pi \int_{0}^{R}T_{0}^{0}r^{2}dr > 0$).

On the other hand, the  boundary conditions require 
continuity of $\nu$, and  $ \nu'$ across the boundary, defined as  $r = R$. 
Consequently, from (\ref{RN}) we get
\begin{equation}
\label{enu}
e^{\nu}= (1 - \frac{M}{R}).
\end{equation} 

\section{The pure-field condition}

The purpose of this section is to construct a model of a charged particle as a 
non-singular concentration of fields. With this aim we now introduce
the``purely gravitational 
field energy", which is
represented by the Weyl tensor.

In a spherically
symmetric space-time all the components of the Weyl tensor are 
proportional 
to the quantity $W$, defined by \cite{{JJ 2},{JJ 3}}
\begin{equation}
\label{W}
W = \frac{r}{6} 
-\frac{r^{3}e^{-\lambda}}{6}( \frac{{\nu}''}{2} +
 \frac{{\nu'}^{2}}{4} - \frac{{{\nu}'}-{{\lambda}'}}{2r} - 
\frac{{\nu}'{\lambda}'}{4} + \frac{1}{r^{2}}).
\end{equation}
Now using the Einstein-Maxwell  equations (\ref{density})-(\ref{tangential pressure}), one can show that 
\begin{equation}
\label{grav mass and Weyl}
M_{G} = [ W + \frac{4\pi r^{3}}{3}(\rho + 2p_{r} + p_{\perp})]
e^{(\nu + \lambda)/2}.
\end{equation}
This expression\footnote{This result is general, in the sense
 that it does not assume (\ref{strong energy condition}). } 
is interesting because it gives the effective mass
 as the sum of two
parts only; $W$ and $(\rho + 2p_{r} + p_{\perp})$, for the ``purely 
gravitational field" and matter contribution, 
respectively. It suggests that the quantity
$(\rho + 2p_{r} + p_{\perp})e^{(\nu + \lambda)/2}$ can be 
interpreted
as a kind of {\em ``average"} effective density\footnote{In the case 
of perfect fluid the term $(\rho + 2p_{r} + p_{\perp})$ 
reduces to the familiar expression
$(\rho + 3p)$.} of the matter 
inside a sphere of 
radius $r$.

Thus,
an extended particle consisting of ``pure-field" 
 is obtained by setting the matter terms equal to 
zero\footnote{Note that in the literature
the condition for the mass to be of electromagnetic origin 
is normally taken as $\rho = 0$. Instead of this, from 
the arguments given 
above, we see that a better condition would be 
(\ref{equation of state}).}    
\begin{equation}
\label{equation of state}
(\rho + 2p_{r} + p_{\perp}) = 0
\end{equation}
The effective gravitational mass arises 
completely from the Weyl tensor, viz.,
\begin{equation}
\label{Schw mass}
M_{G}(r) = e^{(\nu + \lambda)/2} W(r)
\end{equation}
In this sense, the ``equation of state" (\ref{equation of state})
 generates  
a model wherein the particle is composed only of charge and 
gravitational energy and, consequently, can be interpreted as a 
singularity-free concentration of fields.
 
Let us now focus on the properties of the model. Substituting 
(\ref{enu}) and (\ref{equation of state}) into the field equations 
we get
\begin{equation}
\label{elambda} 
 e^{-\lambda}(\frac{1}{r^{2}}\ + \frac{{\lambda}'}
{2r}) - \frac{1}{r^{2}} = 0.
\end{equation}
This equation can be easily integrated as
\begin{equation}
\label{sol elambda}
e^{-\lambda} = 1 + C r^{2},
\end{equation}
where $C$ is a constant of integration, to be defined from the 
boundary conditions.

The final form of the interior metric is then
\begin{equation}
\label{example 1}
ds^{2} = (1 - \frac{M}{R})dt^{2} - 
(1 - \frac{Mr^{2}}{R^{3}})^{-1}dr^{2} 
 - r^{2}(d\theta^{2} + 
sin^{2}\theta d\phi^{2}),
\end{equation}

The corresponding ``equations of state" are 
\begin{equation}
\label{equation for the radial pressure}
\rho = - p_{r} + \frac{M}{4 \pi R^{3}},
\end{equation}
\begin{equation}
\label{equation for the tangential pressure}
\rho = p_{\perp} + \frac{M}{2 \pi R^{3}}.
\end{equation}
Note that $\mid dp_{r}/d\rho \mid = \mid dp_{\perp}/d\rho \mid = 1$. 
Which means that the distribution is  consistent with 
the ``causality condition" $\mid dp/d\rho \mid \leq 1$ (See for 
example Ref. \cite{Adler}).

\section{Uniform charge density}

In order to illustrate our model, we assume that the charge density ${\rho}_{e}(r)$ is constant 
throughout
the sphere. This is equivalent to assuming that the 
proper charge density $\hat{\rho}_{e}$ varies 
as\footnote{This assumption was used by Tiwari, Rao and Kanakamedala
in their study of electromagnetic mass models \cite{TRK}.}
\begin{equation}
\label{constant charge density}
\hat{\rho}_{e}(r) = \hat{\rho}_{e}(0) e^{-\lambda(r)/2},
\end{equation}
where $\hat{\rho}_{e}(0)$ is the constant charge density at 
$r = 0$. 

The final form of the matter distribution inside the charge is 
as follows 
\begin{equation}
\rho(r) = -p_{r}(r) + \frac{M}{4 \pi R^{3}} = 
\frac{3M}{8 \pi R^{3}}(1 -\frac{r^{2}}{3R^{2}}),\;\;\;p_{r}(R) = 0,
\end{equation}
\begin{equation}
E^{2}(r) = \frac{M}{R^{5}} r^{2}.
\end{equation}   

It is not difficult to see that the resulting model is
``physically reasonable", 
in the sense that it is free of 
singularities, $p_{r} = p_{\perp}$ at $r = 0$, 
and $\rho > 0$ as well as 
$\rho \geq \mid p_{i} \mid$ throughout the distribution.
\section{Discussion and conclusions}
We have presented here a general-relativistic version of the old 
Abraham-Lorentz-Poincar$\acute{e}$ model for an extended charged 
particle. 

In contrast to other models in the literature where 
$\rho = 0$, in our 
model the particle contains matter with positive rest density $\rho$
 and positive proper ``inertial" mass 
$4 \pi \int_{0}^{R}\rho r^{2}e^{\lambda/2} dr$.
 Therefore, the matter and charge that make 
up the particle also have a positive 
density, viz., $T^{0}_{0} = (\rho + E^{2}/8 \pi)$. 

The fact that $T_{0}^{0}$ and $\mu$ are different can be 
understood from the following argument. In any volume element there
 is 
not only matter (with positive density) but also certain amount 
of binding energy - which is the 
energy necessary to maintain stability and keep the 
charge together. The effective matter density  
$\mu$ can be interpreted as the sum of the positive density
 $T_{0}^{0}$ and the binding energy, which is negative. Condition
 $\mu_{in}(r) = 0$, in Eq. (\ref{strong energy condition}), 
expresses that
the binding energy, in our model, exactly balances the positive 
contribution 
from the matter and the electrical field. 
Because of this, in our model,  
the effective gravitational mass is nowhere negative. The 
``weak", ``dominant" and ``strong" energy conditions are satisfied and 
{\em there is no gravitational repulsion anywhere}.

The equation of state inside the particle is 
(what we call the pure-field condition) 
$\rho + 2p_{r} + p_{\perp} = 0$, which  is  the anisotropic 
generalization of $(\rho + 3 p) = 0$. This  
condition replaces the 
usual $\rho = 0$ requirement. We note that the equation of 
state $(\rho + 3 p) = 0$ has 
been
considered in different contexts by several authors. Notably, in 
discussions of cosmic strings \cite{Gott and Rees}. Also, it is the 
only equation of state 
consistent with the existence of zero-point fields \cite{Wesson}. 
Another important feature in our model is that the  particle must have 
unequal principal pressures, otherwise it would 
contain some negative rest mass.

From a mathematical viewpoint, 
the pure-field model discussed here (with $\rho + 2p_{r} + p_{\perp} = 0$) 
can be generalized in several ways. For example, one can 
assume $\mu_{in}(r) = const$,  instead of 
$\mu_{in}(r) = 0$ as in (\ref{vanishing density}). In 
this case the ``bare (or intrinsic)  mass" 
of the particle  will be different from zero, i.e.,   
$M_{G}(R) = W(R) \neq 0$,  and consequently $R > Q^2/M$. 
All these models 
share similar properties in the sense that the 
tensions $p_{r} \neq p_{\perp}$ are responsible for 
holding  
the charge together. However, from a physical and historical viewpoint,  they  
are different.  
In the Abraham-Lorentz-Poincar$\acute{e}$ 
model the mass is {\em entirely} of 
electromagnetic origin. While, if we modify 
 (\ref{vanishing density})  this is no longer so. The  
mass of the charged particle is now the sum of 
the bare ``pure gravitational" mass $W(R)$ and the 
electromagnetic mass $Q^2/R$. 

In summary, without the introduction of negative masses, here 
we have been able to construct a simple model where a charged particle 
can be visualized as a 
concentration
 of fields. The positiveness of energy inside the source, as well as the 
energy conditions,
 require the electron to be an ``extreme" Reissner-Nordstr\"{o}m source of 
gravity \cite{PdL2}.

\end{document}